\documentclass[aps,pra,showpacs,floatfix,nofootinbib]{revtex4}
\usepackage{graphicx} 
\usepackage{bm}
\usepackage{amsmath}
\usepackage{amsfonts}
\usepackage{amssymb}%

\usepackage{dcolumn}

\begin{document}

\title{ Post-Wick theorems for  symbolic manipulation of
second-quantized expressions in atomic many-body perturbation theory
}

\author{Andrei Derevianko }
\affiliation{Physics Department, University of
Nevada, Reno, Nevada  89557, USA}

\date{\today}
\begin{abstract}
Manipulating expressions in many-body perturbation theory  becomes  unwieldily with increasing order of the perturbation theory. Here I derive a set of theorems for efficient simplification of such expressions.
The derived rules are specifically designed for implementing with symbolic algebra tools.
As an illustration, we count the numbers of Brueckner-Goldstone diagrams in the first several orders of many-body perturbation theory for matrix elements between two states of a mono-valent system.
\end{abstract}
\maketitle

%
%
Many-body perturbation theory (MBPT) has proven to be a powerful tool
in physics~\cite{FetWal71} and quantum chemistry~\cite{SzaOst82}.
Although MBPT provides a systematic
approach to solving the many-body problem, the number
and complexity of analytical expressions becomes rapidly unwieldily with increasing order of perturbation theory.
At the same time,
exploring higher orders
is desirable for improving accuracy of {\em ab initio}
atomic-structure methods. Here a number of applications may benefit, ranging
from atomic parity violation~\cite{PorBelDer09} and atomic clocks~\cite{BelSafDer06,AngDzuFla06} to a precision characterization
of long-range inter-atomic potentials for ultra-cold collision studies~\cite{DerJohSaf99}.

To overcome an overwhelming complexity of the MBPT in high orders one
has to develop symbolic tools that automate highly repetitive
but error-prone derivation  of many-body diagrams.
The advantage of using symbolic algebra computing for these goals has
been realized for a number of decades.
For example, the pioneering  ``Schoonschip'' program~\cite{VelWil93} and other symbolic packages are employed for evaluating Feynman
diagrams in quantum electrodynamics and high-energy physics. We
also note similar efforts in quantum chemistry~\cite{JanSch91} 
 (see also Ref.~\cite{Hir03} and references therein).
In atomic MBPT, developing symbolic tools was reported by the Notre Dame~\cite{BluGuoJoh87},
Michigan~\cite{PerXiaFlu01}, and very recently by the Sydney~\cite{Dzu09} and Kassel 
groups~\cite{FriManAng08}.

In practical applications of MBPT one
deals with products of strings of creation and annihilation
operators. Typically such products are evaluated with the Wick's theorem (see, e.g., discussion in Ref.~\cite{LinMor86}).
This is point of departure of symbolic calculations described in Refs.~\cite{BluGuoJoh87} and \cite{PerXiaFlu01}.
The application of the Wick's theorem results in a series of Kronecker
delta symbols. The next step in a derivation requires carrying
out summation over the delta symbols. In a typical application the
resulting terms are redundant and require additional efforts
to further simplify and combine the expressions. A complexity of both the
applying the Wick's theorem  and  the further simplification grows
rapidly as the order of perturbation theory increases.

Over the past decade, our group in Reno has developed an alternative set of symbolic
tools for MBPT. The goal of our work was to study high orders of MBPT, e.g.,
fourth-order contributions to matrix elements for mono-valent
atoms~\cite{DerEmm02,CanDer04}. In our practical
work, we found that the conventional approaches based on the
straightforward applications of the Wick's theorem require prohibitively long
computational times.
To overcome this difficulty, I have derived a set of rules enabling efficient
derivation of MBPT expressions
for fermionic systems in high-orders of MBPT. These theorems are reported here.

The Wick's theorem works at the level of elemental pairwise contractions of creation and
annihilation operators.
The basis idea of the present approach is to shortcut directly to
the resulting expressions for a typical operations in MBPT,
without the need to apply expensive pairwise operations.
The theorems are formulated as a set of symbolic replacement rules,
ideally suited for implementing with symbolic algebra systems. We provide an
accompanying Mathematica package downloadable from the author's web-site~\cite{APDweb}.
In this work we focus on mono-valent systems.

%
%

The paper is organized as follows. In Section~\ref{Sec:BackgroundNotation}, we review main
results from the many-body perturbation theory and introduce notation. In Sections~\ref{Sec:OneBody} and \ref{Sec:TwoBody} we derive rules for multiplying second-quantized operators with atomic wave-functions.
Similar theorems are derived for determining MBPT corrections to energies and matrix elements in Section~\ref{Sec:Observables}. Finally, as an illustration, in Section~\ref{Sec:Count} we derive
explicit expressions for matrix elements in several first orders of MBPT and count the number of resulting  diagrams.

\section{Background and notation}
\label{Sec:BackgroundNotation}
\subsection{Second quantization, normal forms, and the Wick's theorems}
We start by recapitulating relevant notation and results from the second-quantization method as applied to fermionic systems.

At the heart of the second quantization technique lies an expansion of the true many-body
wave function over  properly anti-symmetrized products of  single-particle orbitals (the Slater determinants).
The machinery is simplified by introducing the creation ($a^\dagger_k$)
and the annihilation ($a_k$) operators satisfying the anti-commutation relations
\begin{align*}
a_{j}^{\dagger}a_{k}^{\dagger}  &  =-a_{k}^{\dagger}a_{j}^{\dagger},\\
a_{j}a_{k}  &  =-a_{k}a_{j},\\
a_{j}a_{k}^{\dagger}  &  =\delta_{jk}-a_{k}^{\dagger}a_{j} \, ,\\
a_{j}a_{j}  &  \equiv 0 \, ,\\
a_{j}^{\dagger}a_{j}^{\dagger}  &  \equiv 0 \, .
\end{align*}

Applying strings of creation and annihilation operators to the vacuum state $|0\rangle$ builds the
Slater determinants.
A one-particle operator in the second quantization (such as an interaction with external field) reads
\begin{equation}
Z=\sum_{kl} z_{kl} a_{k}^{\dagger}a_{l} \, .
\label{Eq:Z}
\end{equation}
A two-particle operator (such as a pair-wise Coulomb interaction between electrons) is represented as
\begin{eqnarray}
G &=& \frac{1}{2}\sum_{ijkl} g_{ijkl} a_{i}^{\dagger}a_{j}^{\dagger}a_{l}a_{k} \label{Eq:G}\\
&=& \frac{1}{4}\sum_{ijkl} \tilde{g}_{ijkl} a_{i}^{\dagger}a_{j}^{\dagger}a_{l}a_{k}  \label{Eq:Gtilde} \, .
\end{eqnarray}
Matrix elements $z_{kl}$ and $g_{ijkl}$ are conventionally defined on the basis of single-particle orbitals.
Symmetry of the two-particle operator with respect to permuting electron labels leads to $g_{ijkl}=g_{jilk}$.
By renaming summation indexes in Eq.~(\ref{Eq:G}) and
using anti-commutation rules, we may express $G$ in terms of the anti-symmetrized combination
$\tilde{g}_{ijkl} =   {g}_{ijkl} - {g}_{ijlk}$, Eq.~(\ref{Eq:Gtilde}).  Apparently,
swapping indexes leads to the following properties,
\begin{equation}
\tilde{g}_{ijkl} = - \tilde{g}_{ijlk} =  -\tilde{g}_{jikl}= \tilde{g}_{jilk}.
\label{Eq:TildeRulesG}
\end{equation}

Conventionally, in applications of the second quantization technique to many-electron systems,
one distinguishes between three groups of single-particle states (orbitals):
core, virtual (excited) and valence orbitals.
The core orbitals are occupied and form the quasi-vacuum state $|0_c\rangle$. Virtual orbitals
are unoccupied in $|0_c\rangle$. We will treat valence orbitals as a part of the set of virtual
orbitals.
We follow a convention of Ref.~\cite{LinMor86}  and label core
orbitals as $a,b\ldots$, excited (virtual) orbitals as
$m,n,\ldots$, and valence orbitals as $v,w$. Indexes $i,j,k,l$
run over both core and virtual orbitals.
For example, in the independent-particle-approximation, a state of a mono-valent atom
may be represented as $a^\dagger_v |0_c\rangle$, where the quasi-vacuum
state
$|0_{c}\rangle=\left(\prod_{a\in\mathrm{core}}a_{a}^{\dagger }\right)  |0\rangle$
represents a closed-shell atomic core.

Further, we review several results related to the normal form of
operator products, $:\cdots:$. The operators are rearranged so that
$
\mathit{a}_{\mathrm{core}}\mathit{\ and\ a}_{\mathrm{virt}}^{\dagger
}\mathit{\ appear\ to\ the\ left\ of\ a}_{\mathrm{core}}^{\dagger
}\mathit{\ and\ a}_{\mathrm{virt}} \, .
$
When acted on the $|0_{c}\rangle$ quasi-vacuum state, most of the strings of operators in the normal form  produce
vanishing result.

One of the central results related to the normal forms is the  {\em Wick's expansion into normal products}
\[
A=:{A}:+:{\overline{A}}: \, ,
\]
$:{\overline{A}}:$ being a sum of normal ordered terms obtained by making all
possible single, double, triple ... contractions within $A$. Contractions between two creation/annihilation
operators $x$ and $y$ are
defined as $\overline{xy}\equiv xy-:{xy}:$.
The overall sign of $:{A}:$ is $(-1)^{p}$, $p$ being a number of permutations to bring $A$ into the normal form. The same rule holds for terms in $:\overline{A}:$ - we count permutations necessary to
bring the  operators being contracted next to each other and also the permutations needed to arrange the
resulting term in normal form.
The only nonvanishing contractions are
\[
\overline{a_{m}a_{n}^{\dagger}}=\delta_{mn},\;\overline{a_{a}^{\dagger}a_{b}%
}=\delta_{ab}.
\]
All contractions between the core and excited (including valence) orbitals vanish.

Using the Wick's expansion,
one can rewrite a single-particle
operator $Z$, Eq.(\ref{Eq:Z}), as a sum of zero--body (scalar) and one--body contributions~\cite{LinMor86}
\begin{align}
Z & = Z_0 + Z_1 \, , \label{Eq:Znormal}\\
Z_0  &  =\sum_{a} z_{aa}\,,\nonumber \\
Z_1  &  =\sum_{ij}z_{ij}  \,:a_{i} ^{\dagger}a_{j}:\, .\nonumber
\end{align}
Similarly,
any two-particle operator $G$, Eq.(\ref{Eq:G}), may be represented as a sum of zero-body
$G^{(0)}$, one-body $G^{(1)}$ and two-body $G^{(2)}$ terms,
\begin{align}
G & = G_0 + G_1 +G_2\, , \label{Eq:Gnormal} \\
G_0  &  =\frac{1}{2}\,\sum_{ab}\tilde{g}_{abab}\,,\nonumber \\
G_1  &  =\sum_{ij}\left(  \sum_{a}\tilde{b}_{iaja}\right)  \,:a_{i}%
^{\dagger}a_{j}:\,, \nonumber \\
G_2  &  =\frac{1}{4}\,\sum_{ijkl}\tilde{b}_{ijkl}\,:a_{i}^{\dagger}a_{j}^{\dagger
}a_{l}a_{k}:. \nonumber
\end{align}

Technically, the MBPT formalism requires multiplying second-quantized operators.
Simplification of the resulting expressions is greatly aided by  {\em the Wick's theorem for operator products}.
For two strings of operators in the normal form $:A:$ and $:B:$, the theorem states
\begin{equation}
:A: \, :B:=:AB:+ :\overline{AB}: \label{Eq:WickAB} \, .
\end{equation}
Here $:\overline{AB}:$ represents the sum of the normal-ordered terms with all
possible contractions between the operators in $A$ and those in $B$.

\subsection{MBPT for mono-valent atoms}
To gain insight into a general structure of MBPT expressions,
here we briefly reiterate MBPT formalism~\cite{DerEmm02}
for atoms with a single valence electron outside a closed-shell
core.
The lowest-order valence wavefunction is simply
$|\Psi_v^{(0)}\rangle = a^\dagger_v | 0_c \rangle$,  where $v$ is a  valence orbital.
The perturbation
expansion is built in powers of the residual interaction $V_I$ defined as
a difference between the full Coulomb interaction between
the electrons and the model potential used to generate the single-particle basis.
The $n^\mathrm{th}$-order correction to the valence wavefunction may be expressed as
\begin{equation}
|\Psi_v^{(n)}\rangle = -R_{v}\left\{  Q\,V_I\,|\Psi_v^{(n-1)}\rangle\right\}
_{\mathrm{linked}} \, ,
\label{Eq:Psin}
\end{equation}
where $R_v$ is a resolvent operator modified to include so-called ``folded''
diagrams~\cite{DerEmm02}, projection operator $Q=1-|\Psi_v^{(0)}\rangle \langle \Psi_v^{(0)}| $,
and only linked diagrams~\cite{LinMor86} are to be kept.
For mono-valent systems a convenient starting point
is a single-particle basis generated in the frozen-core ($V^{N-1}$)  Hartree-Fock (HF)
approximation~\cite{Kel69}.
In this approximation, the residual interaction is simplified to a
two-body part, $G_2$, of the Coulomb interaction
and  the number of  MBPT diagrams is substantially reduced~\cite{LinMor86,BluGuoJoh87}.

The recursion relation,  Eq.~(\ref{Eq:Psin}), systematically solves the many-body problem,  as we may generate corrections to the wave-function at any
given order of the perturbation theory. With such calculated corrections to wave-functions of two valence
states $w$ and $v$, the $n^\mathrm{th}$-order contribution to matrix
elements of an operator $\hat{Z}$ may be determined as
\begin{equation}
Z^{(n)}_{wv} = \sum_{k=0}^{n-1}
\langle \Psi_w^{(n-k-1)} |  Z  |\Psi_v^{(k)}  \rangle_\mathrm{val,\,conn}
+ Z^{(n)}_{wv, \, {\rm norm}}\label{Eq:Zn}     \, .
\end{equation}
Here $Z^{(n)}_{wv, \, {\rm norm}}$ is a normalization correction arising due to an intermediate
normalization scheme employed in derivation of Eq.~(\ref{Eq:Psin}). Subscript
``$\mathrm{val,\,conn}$'' indicates that only connected diagrams involving excitations
from valence orbitals are included in the expansion.

\subsection{Generic contribution to wavefunction}
Now we would like to introduce short-hand notation for strings of creation ($a^\dagger_k$)
and annihilation ($a_k$) operators in the Fermi statistics. String of $x$ operators
\[
E_{\alpha}^{\dagger}=a_{1}^{\dagger}a_{2}^{\dagger}\cdots a_{x}^{\dagger}
\]
combines creation operators  for excited orbitals and symbol $\alpha$ ranges over the set $1,2\ldots x$.
Similarly,
\[
C_{\beta}=a_{1}a_{2}\cdots a_{y}%
\]
represents a string of $y$ annihilation operators for core orbitals, with
symbol $\beta$ spanning the indexes $1,2\ldots y$. Finally, $V^{\dagger}$ is
either $a_{v}^{\dagger}$ or $1$ depending on the presence of the valence
creation operator.

On very general grounds a generic piece of atomic wave-function for a mono-valent atom may be
represented as
\begin{equation}
|\Phi \rangle=\sum_{\left\{  \alpha\right\}  ,\left\{
\beta\right\}  }L\left[  \left(  1,2\ldots x\right)  _{\alpha},\left(
1,2\ldots y\right)  _{\beta}\right]  \,\,E_{\alpha}^{\dagger}\,C_{\beta
}\,V^{\dagger}\,|0_{c}\rangle \, ,
\label{Eq:PhiGeneric}
\end{equation}
where $\sum_{\left\{  \alpha\right\}  }=\sum_{1}\cdots\sum_{x}$, and
 $L\left[  \left(  1,2\ldots x\right) _{\alpha},\left(
1,2\ldots y\right) _{\beta}\right]$ is a c-number object which
depends on the indexes.

As an illustration, a  doubly-excited core state of a mono-valent atom
may read $\sum_{mnab} \rho_{mnab} a_{m}^\dagger a_{n}^\dagger a_a a_b a^\dagger_v |0_c \rangle$. Apparently, $x=y=2$, $E_{\alpha}^{\dagger}=a_{m}^\dagger a_{n}^\dagger$,
$C_{\beta}=a_a a_b$, $V^{\dagger}=a^\dagger_v$  and the groups of indexes are $\left(  1,2\ldots x\right)  _{\alpha}= (m,n)$ and $\left(1,2\ldots y\right) _{\beta} = (a,b)$. In summations, the symbolic index $\alpha$ would run over symbols $m$ and $n$ (which range over all excited orbitals). Similarly, the symbolic index  $\beta$ assumes symbolic values $a$ and $b$, i.e., the labels of the core orbitals.

The Eq.~(\ref{Eq:PhiGeneric}) will be the central object in our derivations presented below. We will act on this ``generic piece of  wave-function'' with one-- and two--body operators and also bring the resulting expressions into the very same form of Eq.~(\ref{Eq:PhiGeneric}).

\section{Simplification theorems}
We would like to follow the MBPT prescription~(\ref{Eq:Psin}) and compute the wave-function in an arbitrary order.
To this end we will derive rules for simplifying products of  one-- and two--particle operators with the
generic wave-function $|\Phi \rangle$, Eq.(\ref{Eq:PhiGeneric}).
The wave-function $|\Phi \rangle$ is in the normal form with respect to the quasi-vacuum state.
Therefore, we take the second-quantized operators expanded in the normal forms, Eqs.(\ref{Eq:Znormal},\ref{Eq:Gnormal}),  and apply the Wick's theorem~(\ref{Eq:WickAB}). Apparently, the zero--body terms $Z_0$ and $G_0$  (c-numbers) do not produce non-trivial results and in the derivation below we focus on contractions of one-- and two--body operators with strings of operators entering $|\Phi \rangle$.
Finally, we  bring the resulting chain of operators in the same standardized sequence of  operators
$E_{\alpha}^{\dagger}\,C_{\beta}\,V^{\dagger}$, as in (\ref{Eq:PhiGeneric}); of course the number of operators in each group may differ
from the starting numbers of operators in $|\Phi\rangle$.

\subsection{Product with a one-body operator}%
\label{Sec:OneBody}
Here we focus on acting with the operator
\begin{equation}
Z_1=\sum_{ij}z(i,j)\,:a_{i}^{\dagger}a_{j}: \label{Eq:Z1}
\end{equation}
on a generic wave-function  $|\Phi\rangle$, Eq.(\ref{Eq:PhiGeneric}).
 According
to the Wick's theorem, we may have 0, 1, and 2 contractions between $Z_1$ and operators entering  $|\Phi\rangle$.
There are only 6 distinct possibilities
classified by the number and type of contractions
\begin{align*}
\left\{  Z_1\,|\Phi\rangle\right\}&=\left\{  Z_1\,|\Phi\rangle\right\}_0  \\
 &+\left\{  Z_1\,|\Phi\rangle\right\}_{1e} + \left\{  Z_1\,|\Phi\rangle\right\}_{1c}  +\left\{  Z_1\,|\Phi\rangle\right\}_{1v} \\
 &+\left\{  Z_1\,|\Phi\rangle\right\}_{1e,1c} +\left\{  Z_1\,|\Phi\rangle\right\}_{1v,1c} \, .
 \end{align*}
 The first term corresponds to no contractions. The term $1e$ results from contracting one excited orbital from $Z_1$ and one orbital from the string $E_{\alpha}^{\dagger}$ of the operators entering $|\Phi\rangle$.
Subscript  ${1v,1c}$ labels double contraction: one contraction $1v$ involves an operator from the valence string $V^\dagger$ and the other contraction $1c$ involves an operator from the core string $C_\beta$. The labeling scheme for other contributions follows from these examples.

Notice, that a straight-forward application of the Wick theorem~(\ref{Eq:WickAB}) is inefficient. Operators
resulting from expanding Eq.(\ref{Eq:WickAB}) ultimately  act on $|0_c\rangle$ and many terms in the Wick's expansion will produce zero result.
Indeed, we may write explicitly
\[
Z_1=
   \sum_{mn}z(m,n) \, a_{m}^{\dagger}a_{n}
  +\sum_{ma}z(a,m) \, a_{a}^{\dagger}a_{m}
  -\sum_{ab}z(a,b) \, a_{b}a_{a}^{\dagger}
  +\sum_{ma}z(m,a) \, a_{m}^{\dagger}a_{a}\, .
\]
As an example, consider term with no contractions. Only the last contribution from the above expansion will contribute, because the first two terms annihilate an unfilled orbital in $|0_c\rangle$ and the third term promotes an electron into already occupied core orbital. Based on this discussion, we shortcut the application of the Wick theorem and limit ourself to a much smaller subset of terms.

\subsubsection{No contractions }%
In this case, the relevant part of the operator  $Z_1$ contains one core annihilation operator
and one creation operator involving excited orbital.
Then we may simply move the operators from $Z_1$ to the ends of the excited $E_{\alpha}^{\dagger}$ and
core $C_\beta$ operator strings in the wave-function. The additional core and excited orbital
indexes are absorbed in the summation.
\[
\left\{  Z_1\,|\Phi\rangle\right\}  _{0}=\sum_{\left\{
\alpha\right\}  ,\left\{  \beta\right\}  }\,L^{0}\left[  \left(  1,2\ldots
x,x+1\right)  _{\alpha},\left(  1,2\ldots y,y+1\right)  _{\beta}\right]
\,\left(  E_{\alpha}^{\dagger}a_{x+1}^{\dagger}\right)  \,\left(  C_{\beta
}a_{y+1}\right)  \,V^{\dagger}\,|0_{c}\rangle \, ,
\]%
where
\[
L^{0}\left[  \left(  1,2\ldots x,x+1\right)  _{\alpha},\left(  1,2\ldots
y,y+1\right)  _{\beta}\right]  =(-1)^{y}\,z\left(  x+1,y+1\right)  \,L\left[
\left(  1,2\ldots x\right)  _{\alpha},\left(  1,2\ldots y\right)  _{\beta
}\right] \, .
\]
Moving a pair operators together does not produce a phase.
The phase factor appears because the excited orbital operator was additionally moved through (anti-commuted with) $y$
operators in the $C_\beta$ string.

\subsubsection{Single contraction with the creation operator in the $E_{\alpha
}^{\dagger}$ group: 1e}
There are $x$ such contractions ($\,\overline{a_{j}a_{\alpha}^{\dagger}}=\delta_{j\alpha}$); we contract the operators in turn. By contracting with
$a_{1}^{\dagger}$ we obtain a string of operators $a_{2}^{\dagger}\cdots
a_{x}^{\dagger}$ and we bring the $a_{i}^{\dagger}$ to the end of this string thus
acquiring a phase $\left(  -1\right)  ^{x-1}$. By renaming the dummy summation
indexes we may bring the resulting sequence of operators to the same form
$a_{2}^{\dagger}\cdots a_{x}^{\dagger}a_{i}^{\dagger}$. For example, as a
result of contracting with $a_{\mu}^{\dagger}$ we obtain
\[
\delta_{j\mu}\,\left(  -1\right)  ^{\mu-1}\left(  -1\right)  ^{x-1}%
a_{1}^{\dagger}a_{2}^{\dagger}\cdots a_{\mu-1}^{\dagger}a_{\mu-1}^{\dagger
}a_{x}^{\dagger}a_{i}^{\dagger} .
\]
Now we rename $\left(  \mu\longleftrightarrow1\right)  $ and bring the
resulting string into the form $a_{2}^{\dagger}\cdots a_{x}^{\dagger}%
a_{i}^{\dagger}$
\begin{align*}
&  \delta_{j1}\,\left(  -1\right)  ^{\mu-1}\left(  -1\right)  ^{x-1}a_{\mu
}^{\dagger}\,a_{2}^{\dagger}\cdots a_{\mu-1}^{\dagger}a_{\mu-1}^{\dagger}%
a_{x}^{\dagger}a_{i}^{\dagger}=\\
&  \delta_{j1}\,\left(  -1\right)  ^{\mu-1}\left(  -1\right)  ^{x-1}\,\left(
-1\right)  ^{\mu}\,\,a_{2}^{\dagger}\cdots a_{\mu-1}^{\dagger}a_{\mu}%
^{\dagger}a_{\mu-1}^{\dagger}a_{x}^{\dagger}a_{i}^{\dagger}=\\
&  -\delta_{j1}\left(  -1\right)  ^{x-1}\,\,a_{2}^{\dagger}\cdots a_{\mu
-1}^{\dagger}a_{\mu}^{\dagger}a_{\mu-1}^{\dagger}a_{x}^{\dagger}a_{i}%
^{\dagger}%
\, .
\end{align*}

Finally,
\[
\left\{  Z\,|\Phi\rangle\right\}  _{1e}=\sum_{\left\{
\alpha\right\}  ,\left\{  \beta\right\}  }\,L^{1e}\left[  \left(  2\ldots
x,x+1\right)  _{\alpha},\left(  1\ldots y\right)  _{\beta}\right]  \,\left(
a_{2}^{\dagger}\cdots a_{x}^{\dagger}a_{x+1}^{\dagger}\right)  _{\alpha
}\,C_{\beta}\,V^{\dagger}\,|0_{c}\rangle \, ,
\]
with%
\[
L^{1e}\left[  \left(  2\ldots x,x+1\right)  _{\alpha},\left(  1,2\ldots
y\right)  _{\beta}\right]  =\left(  -1\right)  ^{x-1}\sum_{1_{e}}z\left(
x+1,1_{e}\right)  \mathcal{A}_{1_{e}}L\left[  \left(  1_{e},2\ldots x\right)
_{\alpha},\left(  1,\ldots y\right)  _{\beta}\right]  ,
\]
where the operator $\mathcal{A}_{1_{e}}$ anti-symmetrizes $L$ over the first excited index,
i.e.,
\begin{equation}
\mathcal{A}_{1_{e}}L\left[  \left(  1_{e},2\ldots x\right)  _{\alpha},\left(
{}\right)  _{\beta}\right]  =L\left[  \left(  1_{e},2\ldots x\right)
_{\alpha},\left(  {}\right)  _{\beta}\right]  -L\left[  \left(  2,1_{e}%
,3,\ldots x\right)  _{\alpha},\left(  {}\right)  _{\beta}\right]  -L\left[
\left(  3,2,1_{e},\ldots x\right)  _{\alpha},\left(  {}\right)  _{\beta
}\right]  -\ldots
\end{equation}
or%
\begin{equation}
\mathcal{A}_{i}f\left(  1,2..,x\right)  =f\left(  1,2..,x\right)  -\sum
_{\mu\not =i}f\left(  1,2..,\mu,\ldots x\right)  _{\mu\longleftrightarrow i} \, .
\label{Eq:Apartanti1}
\end{equation}
As an example,
\[
\mathcal{A}_{1_{e}} \rho_{mn,ab}  = \rho_{mn,ab}
 -\rho_{nm,ab}\, .
 \]
Computationally, in symbolic algebra implementations, the symbol replacement operations in Eq.(\ref{Eq:Apartanti1})  are efficient.
We note that $\mathcal{A}_{1_{e}}\widetilde{L}=x\,\widetilde{L}$.

\subsubsection{Single contraction with the annihilation operator in the
$C_{\beta}$ group: 1c}

The derivation is similar to the previous case; $\overline{a_{i}^{\dagger
}a_{\beta}}=\delta_{i\beta}.$ There is a phase factor $\left(  -1\right)
^{x+y}$ due to the transfer  of $a_{j}$ through $E_{\alpha}^{\dagger}\,C_{\beta}$ and an
additional factor $\left(  -1\right)  ^{x}$ due to moving the $a_{i}^{\dagger}\,$ to
the beginning of the $C_{\beta}$ group, resulting in the total phase of
$\left(  -1\right)  ^{y}$.
The result reads
\[
\left\{  Z\,|\Phi\rangle\right\}  _{1c}=\sum_{\left\{
\alpha\right\}  ,\left\{  \beta\right\}  }\,L^{1c}\left[  \left(  1\ldots
x\right)  _{\alpha},\left(  2\ldots y,y+1\right)  _{\beta}\right]  E_{\alpha
}^{\dagger}\,\left(  a_{2}\cdots a_{y}a_{y+1}\right)  _{\beta}\,\,V^{\dagger
}\,|0_{c}\rangle \, ,
\]%
\[
\,L^{1c}\left[  \left(  1\ldots x\right)  _{\alpha},\left(  2\ldots
y,y+1\right)  _{\beta}\right]  =\left(  -1\right)  ^{y}\sum_{1_{c}}z\left(
1_{c},y+1\right)  \mathcal{A}_{1_{c}}L\left[  \left(  1\ldots x\right)
_{\alpha},\left(  1_{c},2,\ldots y\right)  _{\beta}\right] \, .
\]
The operator $\mathcal{A}_{1_{c}}$ anti-symmetrizes $L$ over the first core index and
is defined similarly to $\mathcal{A}_{1_{e}}$.

\subsubsection{Single contraction with the valence creation operator: 1v}

Contraction $\,\overline{a_{j}a_{v}^{\dagger}}=\delta_{jv}$. Phase $\left(
-1\right)  ^{x+y}$ arises due to bringing $a_{j}$ to the end of $E_{\alpha}^{\dagger
}\,C_{\beta}$ and the phase $\left(  -1\right)  ^{x}$ due to transferring $a_{i}^{\dagger}$ to
the end of $E_{\alpha}^{\dagger}$. The result reads
\[
\left\{  Z\,|\Phi\rangle\right\}  _{1v}=\sum_{\left\{
\alpha\right\}  ,\left\{  \beta\right\}  }\,L^{1v}\left[  \left(  1\ldots
x,x+1\right)  _{\alpha},\left(  1\ldots y\right)  _{\beta}\right]  \,\left(
a_{1}^{\dagger}\cdots a_{x+1}^{\dagger}\right)  _{\alpha}\,C_{\beta}%
\,\,|0_{c}\rangle ,
\]
with
\[
L^{1v}\left[  \left(  1\ldots x,x+1\right)  _{\alpha},\left(  1,2\ldots
y\right)  _{\beta}\right]  =\left(  -1\right)  ^{y}z\left(  x+1,v\right)
\mathcal{\,}L\left[  \left(  1,\ldots x\right)  _{\alpha},\left(  1,\ldots
y\right)  _{\beta}\right]  .
\]

\subsubsection{Double contractions: one excited and one core operators (1e,1c)}%
Here the number of operators in both core and excited orbitals strings is reduced by one.
The derivation is similar to the $1e$ case.
\[
\left\{  Z\,|\Phi\rangle\right\}  _{1e,1c}=\sum_{\left\{
\alpha\right\}  ,\left\{  \beta\right\}  }\,L^{1e,1c}\left[  \left(  2,\ldots
x\right)  _{\alpha},\left(  2,\ldots y\right)  _{\beta}\right]  \,\left(
a_{2}^{\dagger}\cdots a_{x}^{\dagger}\right)  _{\alpha}\,\left(  a_{2}\cdots
a_{y}\right)  _{\beta}\,V^{\dagger}\,\,\,|0_{c}\rangle \, ,
\]

\[
L^{1e,1c}\left[  \left(  2,\ldots x\right)  _{\alpha},\left(  2,\ldots
y\right)  _{\beta}\right]  =\left(  -1\right)  ^{x-1}\sum_{1_{c},1_{e}%
}z\left(  1_{c},1_{e}\right)  \mathcal{A}_{1_{c}}\mathcal{A}_{1_{e}}L\left[
\left(  1_{e}\ldots x\right)  _{\alpha},\left(  1_{c},\ldots y\right)
_{\beta}\right]\, .
\]

\subsubsection{Double contractions : valence and one core operators (1v,1c)}%

\[
\left\{  Z\,|\Phi\rangle\right\}  _{1v,1c}=\sum_{\left\{
\alpha\right\}  ,\left\{  \beta\right\}  }\,L^{1v,1c}\left[  \left(  1,\ldots
x\right)  _{\alpha},\left(  2,\ldots y\right)  _{\beta}\right]  E_{\alpha
}^{\dagger}\left(  a_{2}\cdots a_{y}\right)  _{\beta}\,\,\,|0_{c}\rangle \, ,
\]%
\[
L^{1v,1c}\left[  \left(  1,\ldots x\right)  _{\alpha},\left(  2,\ldots
y\right)  _{\beta}\right]  =\left(  -1\right)  ^{y}\sum_{1_{c}}z\left(
1_{c},v\right)  \mathcal{A}_{1_{c}}L\left[  \left(  1\ldots x\right)
_{\alpha},\left(  1_{c},2,\ldots y\right)  _{\beta}\right] \, .
\]

\subsection{Example: zeroth-order Hamiltonian}
To illustrate the derived simplification rules consider a zeroth-order Hamiltonian
in the second quantization,
\[
H_{0}=\sum_{i}\varepsilon_{i}:a_{i}^{\dagger}a_{i}: \, .
\]
Apparently, this is a special case of a one-body operator, Eq.(\ref{Eq:Z1}), with $z(i,j) \equiv \varepsilon_{i}\,\delta_{ij}$.
While the result of applying $H_0$ to $\Phi$ is trivial, arriving at it via application of the derived rules is instructive.
We would like to show that
\begin{equation}
H_{0}\,|\Phi\rangle=\left(  \Sigma_{\alpha}\varepsilon_{\alpha}-\Sigma_{\beta
\,}\varepsilon_{\beta}+\delta_{v}\,\varepsilon_{v}\right) |\Phi\rangle \, .
\label{Eq:H0xPsin}
\end{equation}
Here $\delta_{v}=0$ if there no valence operator
present in $|\Phi\rangle$ and $\delta_{v}=1$ otherwise. $\Sigma_{\alpha}$ is a sum over all indexes in the $E_{\alpha
}^{\dagger}$ string and  $\Sigma_{\beta}$ is a sum over core indexes in the $C_\beta$ group.

Because the matrix element $z(i,j)=\varepsilon_{i}\,\delta_{ij}$ is diagonal, the only contraction classes which
contribute are $1e$, $1c$, and $1v$.
For example, consider the case of $1e$. We deal with the object
\begin{align*}
L^{1e}\left[  \left(  2\ldots x,x+1\right)  _{\alpha},\left(  1,2\ldots
y\right)  _{\beta}\right]   &  =\left(  -1\right)  ^{x-1}\sum_{1_{e}%
}\varepsilon_{x+1}\,\delta_{1_{e},x+1}\mathcal{A}_{1_{e}}L\left[  \left(
1_{e},2\ldots x\right)  _{\alpha},\left(  1,\ldots y\right)  _{\beta}\right]
\\
&  =\left(  -1\right)  ^{x-1}\varepsilon_{x+1}\mathcal{A}_{x+1}L\left[
\left(  x+1,2\ldots x\right)  _{\alpha},\left(  1,\ldots y\right)  _{\beta
}\right] \, .
\end{align*}%
Then
\[
\left\{  H_{0}\,|\Phi\rangle\right\}  _{1e}=\left(
-1\right)  ^{x-1}\sum_{\left\{  \alpha\right\}  }\sum_{\left\{  \beta\right\}
}\varepsilon_{x+1}\mathcal{A}_{x+1}L\left[  \left(  x+1,2\ldots x\right)
_{\alpha},\left(  1,\ldots y\right)  _{\beta}\right]  \left(  a_{2}^{\dagger
}\cdots a_{x}^{\dagger}a_{x+1}^{\dagger}\right)  _{\alpha}\,C_{\beta
}\,V^{\dagger}\,|0_{c}\rangle \, .
\]
Further rename $x+1\rightarrow1$ and place the $a_{1}^{\dagger}$ at the beginning of the
string:
\[
\left\{  H_{0}\,|\Phi\rangle\right\}  _{1e}%
=\sum_{\left\{  \alpha\right\}  }\sum_{\left\{  \beta\right\}  }%
\varepsilon_{1_{e}}\mathcal{A}_{1_{e}}L\left[  \left(  1_{e},2\ldots x\right)
_{\alpha},\left(  1,\ldots y\right)  _{\beta}\right]  \left(  a_{1_{e}%
}^{\dagger}\cdots a_{x}^{\dagger}\right)  _{\alpha}\,C_{\beta}\,V^{\dagger
}\,|0_{c}\rangle \, .
\]
On expanding the partial anti-symmetrization we encounter  terms
\[
\sum_{\left\{  \alpha\right\}  }\sum_{\left\{  \beta\right\}  }-\varepsilon
_{1_{e}}L\left[  \left(  \mu,2,..,1_{e},..x\right)  _{\alpha},\left(  1,\ldots
y\right)  _{\beta}\right]  \left(  a_{1_{e}}^{\dagger}\cdots a_{x}^{\dagger
}\right)  _{\alpha}\,C_{\beta}\,V^{\dagger}\,|0_{c}\rangle \, .
\]
Rename $1_{e}\longleftrightarrow\mu$ and swap the operators%
\[
\sum_{\left\{  \alpha\right\}  }\sum_{\left\{  \beta\right\}  }+\varepsilon
_{\mu}L\left[  \left(  1_{e},2,..,\mu,..x\right)  _{\alpha},\left(  1,\ldots
y\right)  _{\beta}\right]  \left(  a_{1_{e}}^{\dagger}\cdots a_{x}^{\dagger
}\right)  _{\alpha}\,C_{\beta}\,V^{\dagger}\,|0_{c}\rangle \, .
\]
Therefore,
\[
\left\{  H_{0}\,|\Phi\rangle\right\}  _{1e}%
=\sum_{\left\{  \alpha\right\}  ,\left\{  \beta\right\}  }\left(  \Sigma_{\alpha
}\,\varepsilon_{\alpha}\right)  L\left[  \left(  1,2\ldots x\right)  _{\alpha
},\left(  1,2\ldots y\right)  _{\beta}\right]  \,\,E_{\alpha}^{\dagger
}\,C_{\beta}\,V^{\dagger}\,|0_{c}\rangle \, .
\]
Here $\Sigma_{\alpha}$ is a sum over symbols. Similarly,
\[
\left\{  H_{0}\,|\Phi\rangle\right\}  _{1c}%
=-\sum_{\left\{  \alpha\right\}  ,\left\{  \beta\right\}  }\left(  \Sigma_{\beta
\,}\varepsilon_{\beta}\right)  L\left[  \left(  1,2\ldots x\right)  _{\alpha
},\left(  1,2\ldots y\right)  _{\beta}\right]  \,\,E_{\alpha}^{\dagger
}\,C_{\beta}\,V^{\dagger}\,|0_{c}\rangle \, ,
\]
and$\left\{  H_{0}\,|\Phi\rangle\right\}  _{1v}%
=\varepsilon_{v}|\Phi\rangle$. All the remaining
contractions vanish because they involve matrix elements between core and
excited states. Finally, by adding the derived terms we arrive at the well-known formula (\ref{Eq:H0xPsin}).

\subsection{Contractions with two-body operator}
\label{Sec:TwoBody}
Now we consider products of the two-body part of a two-particle operator $G$ with our
generic piece of the many-body wave function, Eq.(\ref{Eq:Psin}). We will use
\[
G_{2}=\frac{1}{4}\sum_{ijkl}\tilde{g}(i,j,k,l):a_{i}^{\dagger}a_{j}^{\dagger
}a_{l}a_{k}: \, .
\]
The derivation is similar to the one-body case of the preceding Section. Here, however, the maximum number of possible contractions is four and there are 15 distinct cases, enumerated below.

\subsubsection{No contractions}%

\[
\left\{  G_{2}\,|\Phi\rangle\right\}  _{0}=\sum_{\left\{
\alpha\right\}  ,\left\{  \beta\right\}  }\,L^{0}\left[  \left(  1,\ldots
x+2\right)  _{\alpha},\left(  1,\ldots y+2\right)  _{\beta}\right]  \,\left(
E^{\dagger}a_{x+1}^{\dagger}a_{x+2}^{\dagger}\right)  _{\alpha}\,\left(
Ca_{y+1}a_{y+2}\right)  _{\beta}\,V^{\dagger}\,|0_{c}\rangle \, ,
\]%
\[
L^{0}\left[  \left(  1,\ldots x+2\right)  ,\left(  1,\ldots y+2\right)
\right]  =\frac{1}{2}g\left(  x+1,x+2,y+2,y+1\right)  \,L\left[  \left(
1,\ldots x\right)  ,\left(  1,\ldots y\right)  \right]\, .
\]

\subsubsection{Single 1e}%

\[
\left\{  G_{2}\,|\Phi\rangle\right\}  _{1e}%
=\sum_{\left\{  \alpha\right\}  ,\left\{  \beta\right\}  }\,L^{1e}\left[
\left(  2,\ldots x+2\right)  _{\alpha},\left(  1,\ldots y+1\right)  _{\beta
}\right]  \,\left(  a_{2}^{\dagger}\ldots a_{x+1}^{\dagger}a_{x+2}^{\dagger
}\right)  _{\alpha}\,\left(  C\,a_{y+1}\right)  _{\beta}\,V^{\dagger}%
\,|0_{c}\rangle,
\]%
\[
L^{1e}\left[  \left(  2,\ldots x+2\right)  ,\left(  1,\ldots y+1\right)
\right]  =\left(  -1\right)  ^{x+y-1}\frac{1}{2}\sum_{1_{e}}\tilde{g}\left(
x+1,x+2,1_{e},y+1\right)  \mathcal{A}_{1_{e}}\,L\left[  \left(  1_{e},\ldots
x\right)  ,\left(  1,\ldots y\right)  \right]\, .
\]
The partial anti-symmetrization operator $\mathcal{A}_{1_{e}}$ is given by Eq.~(\ref{Eq:Apartanti1}).

\subsubsection{Single 1c}
\[
\left\{  G_{2}\,|\Phi\rangle\right\}  _{1c}%
=\sum_{\left\{  \alpha\right\}  ,\left\{  \beta\right\}  }\,L^{1c}\left[
\left(  1,\ldots x+2\right)  _{\alpha},\left(  2,\ldots y+2\right)  _{\beta
}\right]  \,\left(  E_{\alpha}^{\dagger}a_{x+1}^{\dagger}\right)  _{\alpha
}\,\left(  a_{2}\ldots\,a_{y+1}a_{y+2}\right)  _{\beta}\,V^{\dagger}%
\,|0_{c}\rangle,
\]%
\[
L^{1c}\left[  \left(  1,\ldots x+2\right)  _{\alpha},\left(  2,\ldots
y+2\right)  _{\beta}\right]  =\frac{1}{2}\sum_{1_{c}}\tilde{g}\left(
x+1,1_{c},y+2,y+1\right)  \mathcal{A}_{1_{c}}\,L\left[  \left(  1,\ldots
x\right)  ,\left(  1_{c},\ldots y\right)  \right]\, .
\]

\subsubsection{Single 1v}%

\[
\left\{  G_{2}\,|\Phi\rangle\right\}  _{1v}%
=\sum_{\left\{  \alpha\right\}  ,\left\{  \beta\right\}  }\,L^{1v}\left[
\left(  1,\ldots x+2\right)  _{\alpha},\left(  1,\ldots y+1\right)  _{\beta
}\right]  \,\left(  E_{\alpha}^{\dagger}a_{x+1}^{\dagger}a_{x+2}^{\dagger
}\right)  _{\alpha}\,\left(  C\,a_{y+1}\right)  _{\beta}\,|0_{c}\rangle,
\]%
\[
L^{1v}\left[  \left(  1,\ldots x+2\right)  _{\alpha},\left(  1,\ldots
y+1\right)  _{\beta}\right]  =\frac{1}{2}\tilde{g}\left(
x+1,x+2,v,y+1\right)  \,L\left[  \left(  1,\ldots x\right)  ,\left(  1,\ldots
y\right)  \right]\, .
\]

\subsubsection{Double 2e}%

\[
\left\{  G_{2}\,|\Phi\rangle\right\}  _{2e}%
=\sum_{\left\{  \alpha\right\}  ,\left\{  \beta\right\}  }\,L^{2e}\left[
\ldots\right]  \,\left(  a_{3}^{\dagger}\ldots a_{x+1}^{\dagger}%
a_{x+2}^{\dagger}\right)  _{\alpha}\,C_{\beta}\,V^{\dagger}|0_{c}\rangle,
\]%
\[
L^{2e}\left[  \ldots\right]  =\frac{1}{2}\sum_{1_{e},2_{e}}\tilde{g}\left(
x+1,x+2,1_{e},2_{e}\right)  \mathcal{A}_{1_{e}2_{e}}L\left[  \left(
1_{e},2_{e}\ldots x\right)  ,\left(  1,\ldots y\right)  \right]\, .
\]

The partial anti-symmetrization operator is defined as
\begin{equation}
\mathcal{A}_{1,2}\,f\left(  1,2..x\right)  =f-\sum_{\eta>2}f\left(
1,2,..\eta,..,x\right)  _{\eta\longleftrightarrow 1}-\sum_{\eta>2}f\left(
1,2,..\eta,..,x\right)  _{\eta\longleftrightarrow 2}+\sum_{\eta>\nu>2}f\left(
1,2,..\nu,..\eta,..,x\right)  _{\nu\longleftrightarrow 1,\eta
\longleftrightarrow 2}%
\label{Eq:Apartanti2}
\end{equation}
(this anti-symmetrization produces $x(x-1)/2$ terms).

As an example, $\mathcal{A}_{1_{e}2_{e}}\rho_{mn}=\rho_{mn}$ and $\mathcal{A}_{1_{e}2_{e}} \rho_{mnr}=\rho_{mnr}-\rho_{rnm}$.
Alternatively, this definition may be rewritten as
\[
\mathcal{A}_{1,2}\,f\left(  1,2..x\right)  =\mathcal{A}_{1\,}f+\mathcal{A}%
_{2}\,f-f+2\,f(2,1,...x)+\sum_{\eta>\nu\geq3}f\left(  1,2,..\nu,..\eta
,..,x\right)  _{\nu\longleftrightarrow1,\eta\longleftrightarrow2}
\, .
\]

\subsubsection{Double 2c}

\[
\left\{  G_{2}\,|\Phi\rangle\right\}  _{2c}%
=\sum_{\left\{  \alpha\right\}  ,\left\{  \beta\right\}  }\,L^{2c}\left[
\ldots\right]  \,\left(  E^{\dagger}\right)  _{\alpha}\,\left(  a_{3}\ldots
a_{y+1}a_{y+2}\right)  _{\beta}\,V^{\dagger}|0_{c}\rangle,
\]%
\[
L^{2c}\left[  \ldots\right]  =\frac{1}{2}\sum_{1_{c},2_{c}}\tilde{g}\left(
2_{c},1_{c},y+2,y+1\right)  \mathcal{A}_{1_{c}2_{c}}L\left[  \left(  1,\ldots
x\right)  ,\left(  1_{c},2_{c}\ldots y\right)  \right] \, ,
\]
where the  partial anti-symmetrization operator $\mathcal{A}_{1_{c}2_{c}}$ is defined similarly to
$\mathcal{A}_{1_{e}2_{e}}$.

\subsubsection{Double 1e,1c}

\[
\left\{  G_{2}\,|\Phi\rangle\right\}  _{1e,1c}%
=\sum_{\left\{  \alpha\right\}  ,\left\{  \beta\right\}  }\,L^{1e,1c}\left[
\ldots\right]  \,\,\left(  a_{2}^{\dagger}\ldots a_{x+1}^{\dagger}\right)
_{\alpha}\,\left(  a_{2}\ldots a_{y+1}\right)  _{\beta}\,V^{\dagger}%
|0_{c}\rangle,
\]%
\[
L^{1e,1c}\left[  \ldots\right]  =\left(  -1\right)  ^{x+y+1}\sum_{1_{c},1_{e}%
}\tilde{g}\left(  x+1,1_{c},1_{e},y+1\right)  \mathcal{A}_{1_{c}}%
\mathcal{A}_{1_{e}}L\left[  \left(  1_{e},\ldots x\right)  ,\left(
1_{c},\ldots y\right)  \right].
\]
Here we deal with two successive applications of the partial anti-symmetrization operators introduced in Section~\ref{Sec:OneBody}, Eq.(\ref{Eq:Apartanti1}).

\subsubsection{Double 1c,1v}%

\[
\left\{  G_{2}\,|\Phi\rangle\right\}  _{1v,1c}%
=\sum_{\left\{  \alpha\right\}  ,\left\{  \beta\right\}  }\,L^{1v,1c}\left[
\ldots\right]  \,\,\left(  E^{\dagger}a_{x+1}^{\dagger}\right)  _{\alpha
}\,\left(  a_{2}\ldots a_{y+1}\right)  _{\beta}|0_{c}\rangle,
\]%
\[
L^{1v,1c}\left[  \ldots\right]  =\sum_{1_{c},1_{e}}\tilde{g}\left(
x+1,1_{c},v,y+1\right)  \mathcal{A}_{1_{c}}L\left[  \left(  1,\ldots x\right)
,\left(  1_{c},\ldots y\right)  \right].
\]

\subsubsection{Double 1e,1v}%

\[
\left\{  G_{2}\,|\Phi\rangle\right\}  _{1e,1v}%
=\sum_{\left\{  \alpha\right\}  ,\left\{  \beta\right\}  }\,L^{1e,1v}\left[
\ldots\right]  \,\,\left(  a_{2}^{\dagger}\ldots a_{x+1}^{\dagger}%
a_{x+2}^{\dagger}\right)  _{\alpha}\,C_{\beta}|0_{c}\rangle,
\]%
\[
L^{1e,1v}\left[  \ldots\right]  =\left(  -1\right)  ^{x+y}\sum_{1_{e}}g\left(
x+1,x+2,v,1_{e}\right)  \mathcal{A}_{1_{e}}L\left[  \left(  1_{e},\ldots
x\right)  ,\left(  1,\ldots y\right)  \right].
\]

\subsubsection{Triple 2e,1c}%

\[
\left\{  G_{2}\,|\Phi\rangle\right\}  _{2e,1c}%
=\sum_{\left\{  \alpha\right\}  ,\left\{  \beta\right\}  }\,L^{2e,1c}\left[
\ldots\right]  \,\,\left(  a_{3}^{\dagger}\ldots a_{x+1}^{\dagger}\right)
_{\alpha}\,\left(  a_{2}\ldots a_{y}\right)  _{\beta}\,V^{\dagger}%
|0_{c}\rangle,
\]%
\[
L^{2e,1c}\left[  \ldots\right]  =\sum_{1_{c},1_{e},2_{e}}\tilde{g}\left(
x+1,1_{c},1_{e},2_{e}\right)  \mathcal{A}_{1_{c}}\mathcal{A}_{1_{e},2_{e}%
}L\left[  \left(  1,\ldots x\right)  ,\left(  1_{c},\ldots y\right)  \right].
\]

\subsubsection{Triple 1e,2c}%

\[
\left\{  G_{2}\,|\Phi\rangle\right\}  _{1e,2c}%
=\sum_{\left\{  \alpha\right\}  ,\left\{  \beta\right\}  }\,L^{1e,2c}\left[
\ldots\right]  \,\,\left(  a_{2}^{\dagger}\ldots a_{x}^{\dagger}\right)
_{\alpha}\,\left(  a_{3}\ldots a_{y+1}\right)  _{\beta}\,V^{\dagger}%
|0_{c}\rangle,
\]%
\[
L^{1e,2c}\left[  \ldots\right]  =\left(  -1\right)  ^{x+y+1}\sum_{1_{e}%
,1_{c},2_{c}}\tilde{g}\left(  2_{c},1_{c},1_{e},y+1\right)  \mathcal{A}%
_{1_{e}}\mathcal{A}_{1_{c},2_{c}}L\left[  \left(  1,\ldots x\right)  ,\left(
1_{c},\ldots y\right)  \right].
\]

\subsubsection{Triple 1v,2c}
\[
\left\{  G_{2}\,|\Phi\rangle\right\}  _{1v,2c}%
=\sum_{\left\{  \alpha\right\}  ,\left\{  \beta\right\}  }\,L^{1v,2c}\left[
\ldots\right]  \,\,\left(  E^{\dagger}\right)  _{\alpha}\,\left(  a_{3}\ldots
a_{y+1}\right)  _{\beta}\,|0_{c}\rangle,
\]%
\[
L^{1v,2c}\left[  \ldots\right]  =\sum_{1_{c},2_{c}}\tilde{g}\left(
2_{c},1_{c},v,y+1\right)  \mathcal{A}_{1_{c},2_{c}}L\left[  \left(  1,\ldots
x\right)  ,\left(  1_{c},\ldots y\right)  \right].
\]

\subsubsection{Triple 1e,1v,1c}
\[
\left\{  G_{2}\,|\Phi\rangle\right\}  _{1v,1c,1e}%
=\sum_{\left\{  \alpha\right\}  ,\left\{  \beta\right\}  }\,L^{1v,1c,1e}%
\left[  \ldots\right]  \,\,\left(  a_{2}^{\dagger}\ldots a_{x+1}^{\dagger
}\right)  _{\alpha}\,\left(  a_{2}\ldots a_{y}\right)  _{\beta}\,|0_{c}\rangle,
\]%
\[
L^{1v,1c,1e}\left[  \ldots\right]  =\left(  -1\right)  ^{x+y+1}\sum
_{1_{c},1_{e}}\tilde{g}\left(  x+1,1_{c},1_{e},v\right)  \mathcal{A}_{1_{c}%
}\mathcal{A}_{1_{e}}L\left[  \left(  1,\ldots x\right)  ,\left(  1_{c},\ldots
y\right)  \right].
\]

\subsubsection{Quadruple 2e,2c}%
\[
\left\{  G_{2}\,|\Phi\rangle\right\}  _{2e,2c}%
=\sum_{\left\{  \alpha\right\}  ,\left\{  \beta\right\}  }\,L^{2e,2c}\left[
\ldots\right]  \,\,\left(  a_{3}^{\dagger}\ldots a_{x}^{\dagger}\right)
_{\alpha}\,\left(  a_{3}\ldots a_{y}\right)  _{\beta}\,V^{\dagger}%
|0_{c}\rangle,
\]%
\[
L^{2e,2c}\left[  \ldots\right]  =\sum_{1_{e},2_{e},1_{c},2_{c}}\tilde
{g}\left(  2_{c},1_{c},1_{e},2_{e}\right)  \mathcal{A}_{1_{e},2_{e}%
}\mathcal{A}_{1_{c},2_{c}}L\left[  \left(  1,\ldots x\right)  ,\left(
1_{c},\ldots y\right)  \right].
\]

\subsubsection{Quadruple 1e,1v,2c}%
\[
\left\{  G_{2}\,|\Phi\rangle\right\}  _{1v,1e,2c}%
=\sum_{\left\{  \alpha\right\}  ,\left\{  \beta\right\}  }\,L^{1v,1e,2c}%
\left[  \ldots\right]  \,\,\left(  a_{2}^{\dagger}\ldots a_{x}^{\dagger
}\right)_{\alpha}\,\left(  a_{3}\ldots a_{y}\right)  _{\beta}\,|0_{c}\rangle,
\]%
\[
L^{2e,2c}\left[  \ldots\right]  =\left(  -1\right)  ^{x+y+1}\sum_{1_{e}%
,1_{c},2_{c}}\tilde{g}\left(  2_{c},1_{c},1_{e},v\right)  \mathcal{A}_{1_{e}%
}\mathcal{A}_{1_{c},2_{c}}L\left[  \left(  1,\ldots x\right)  ,\left(
1_{c},\ldots y\right)  \right].
\]

\subsection{Additional remarks}

Note that the introduced partial anti-symmetrization operators, Eq.(\ref{Eq:Apartanti1},\ref{Eq:Apartanti2}) are subsumed into the following general definition of a partial anti-symmetrization operator $\mathcal{A}\left[  \{1,\xi\}_{\gamma},\{1,\mu\}_{\delta}\right]$
over a subset $\{1,\xi\}$
of excited indexes and a subset $\{1,\mu\}\,$\ of core indexes
\begin{align*}
&  \mathcal{A}\left[  \{1,\xi\}_{\gamma},\{1,\mu\}_{\delta}\right]  L\left[
\left(  1,2\ldots x\right)  _{\alpha},\left(  1,2\ldots y\right)  _{\beta
}\right]  =\\
&  \sum\sum\left\{  \digamma\left[  \left(  1,2\ldots x\right)  _{\alpha
}\right]  \,\digamma\left[  \left(  1,2\ldots y\right)  _{\beta}\right]
\,L\left[  \left(  1,2\ldots x\right)  _{\alpha},\left(  1,2\ldots y\right)
_{\beta}\right]  \right\}  _{\left(  \gamma_{i}\longleftrightarrow\alpha
_{j},\delta_{i}\longleftrightarrow\beta_{j}\right)  } \, .
\end{align*}
Here the summation is over all possible renaming of indexes in the groups.
Individual phases of the terms are determined by functions $\digamma$ which we use to denote the conventionally-defined signature of the resulting permutation of indexes.

The derived rules may be presented in a more symmetric form by noticing that
no matter how simple or complicated the dependence on the
indexes inside the object $L$ is, by systematically swapping the dummy summation indexes
the generic piece of MBPT wave-function, Eq.~(\ref{Eq:PhiGeneric}), may be rewritten as
\[
|\Phi\rangle=\frac{1}{x!y!}\sum_{\left\{  \alpha\right\}
,\left\{  \beta\right\}  }\widetilde{L}\left[  \left(  1,2\ldots x\right)
_{\alpha},\left(  1,2\ldots y\right)  _{\beta}\right]  \,\,E_{\alpha}%
^{\dagger}\,C_{\beta}\,V^{\dagger}\,|0_{c}\rangle\,,
\]
where the object $\widetilde{L}$ was obtained by a complete anti-symmetrization of
$L$ inside of the groups of excited and core indexes.
In other words, in the derived theorems one could simply replace
\[
  L\left[ \ldots \right] \rightarrow \frac{1}{x!y!} \widetilde{L}\left[ \ldots \right]
\]
and unfold  the partial anti-symmetrization operators.
 While the resulting expressions may be
more aesthetically appealing, we did not find any particular advantage of using them in practical calculations.

\section{Observables}
\label{Sec:Observables}
Application of the derived rules allows us to find many-body correction to
atomic wave-function in an arbitrary order of MBPT via Eq.~(\ref{Eq:Psin}).
Below we focus on an efficient symbolic evaluation of expressions for MBPT correction
to energies and matrix elements.

\subsection{Corrections to energy}
The correlation correction to energy in the $n$-th order of MBPT may be found with the
$n-1$-th order correction to the wave-function
\begin{equation}
\delta E^{(n)}=\langle\Psi^{(0)}|V_I|\Psi^{(n-1)}\rangle\, ,
\end{equation}
or in the frozen-core approximation for mono-valent atoms,
\begin{equation}
\delta E^{(n)}=\langle0_{c}|a_{v}\,\left\{  G_{2}|\Psi_v^{(n-1)}\rangle\right\}  \,.
\end{equation}
Now we focus on the object in the curly brackets, $|\phi \rangle =  G_{2}|\Psi_v^{(n-1)}\rangle$.
We may use the results of Section~\ref{Sec:TwoBody} and derive a multitude of
the terms on the r.h.s. Ultimately determination of the energy correction is simplified by noticing that only a small number
of terms would remain after forming the required product $\langle 0_{c}|a_{v}|\phi \rangle$.
The
non-vanishing contributions arise from generic pieces
\[
\left\{  G_{2}|\Psi^{(n-1)}\rangle\right\}_{v}=L_{c}[()()] \,a_{v}^{\dagger}%
|0_{c}\rangle,
\]
where $L_{c}$ does not depend on the valence orbital, and
\[
\left\{  G_{2}|\Psi^{(n-1)}\rangle\right\}  _{e}=\sum_{1_{e}}L_{v}%
[(1_{e})()]\,a_{1_{e}}^{\dagger}|0_{c}\rangle,
\]
where $L_v$ necessarily depends on the valence index.
Accordingly, the corrections to the energy may be separated into the core and
valence parts,
\[
\delta E^{(n)}=\delta E_{c}^{(n)}+\delta E_{v}^{(n)},
\]
with $\delta E_{c}^{(n)}=L_{c}$ and $\delta E_{v}^{(n)}=L_{v}[(v)()]$. This solves the problem of
finding the energy correction.

\subsection{Matrix elements of a one-body operator $Z$}
Suppose we derived the MBPT corrections to wave-functions of two valence states $w$ and $v$.
We would like to compute the matrix element of some one-particle operator $Z$. To this end we need
to use the formula (\ref{Eq:Zn}) and compute terms
\[
\langle \Psi_w^{(n-k-1)} |  Z_1  |\Psi_v^{(k)}  \rangle \, .
\]
We start by introducing an intermediate state
\[
|\psi_v \rangle=Z_1 |\Psi_v^{(k)}  \rangle \, .
\]
Then our task will be accomplished by forming the product $\langle \Psi_w^{(n-k-1)}|\psi_v \rangle$.

$|\psi_v \rangle$ may be computed using rules of Section~\ref{Sec:OneBody}.
It will contain a linear combination of various ``generic contributions'';
we focus on a scalar product of generic contributions to $|\psi_v \rangle$
\[
|\Phi_{v}\rangle=\,\sum_{\left\{  \alpha\right\}  ,\left\{  \beta\right\}
}L\left[  \left(  1,2\ldots x\right)  _{\alpha},\left(  1,2\ldots y\right)
_{\beta}\right]  \,\,E_{\alpha}^{\dagger}\,C_{\beta}\,V^{\dagger}%
\,|0_{c}\rangle
\]
and to  $\langle \Psi_w^{(n-k-1)}|$
\[
\langle\Phi_{w}|=\,\sum_{\left\{  \alpha^{\prime}\right\}  ,\left\{
\beta^{\prime}\right\}  }K\left[  \left(  1,2\ldots x^{\prime}\right)
_{\alpha^{\prime}},\left(  1,2\ldots y^{\prime}\right)  _{\beta^{\prime}%
}\right]  \,\langle0_{c}|\,WC_{\beta^{\prime}}^{\dagger}\,E_{\alpha^{\prime}%
}\, .
\]
We adopt a convention that in the groups of operators $C_{\beta^{\prime}%
}^{\dagger}\,$\ and $E_{\alpha^{\prime}}$ the enumeration of operator indexes
goes from right to left, i. e., $C_{\beta^{\prime}}^{\dagger}\,=a_{y^{\prime}%
}^{\dagger}\cdots a_{2^{\prime}}^{\dagger}a_{1^{\prime}}^{\dagger}$ and
$E_{\alpha^{\prime}}=a_{x^{\prime}}\cdots a_{2^{\prime}}a_{1^{\prime}}$.

To obtain a nonzero value for the scalar product $\langle\Phi_{w}|\Phi_{v}\rangle$
we have to perform a maximum number of contractions. Moreover the numbers of operators must be related as
$x+\delta_{v}=x^{\prime}+\delta_{w}$ and $y=y^{\prime}$.

We contract the core orbitals first. Bringing the two groups of operators together
introduces a phase factor of $\left(  -1\right)  ^{\left(  x-x^{\prime
}\right)  \,y}$ and we obtain%
\begin{align*}
\langle\Phi_{w}|\Phi_{v}\rangle &  =\left(  -1\right)  ^{\left(  x-x^{\prime
}\right)  \,y}\sum_{\left\{  \alpha^{\prime}\right\}  ,\left\{  \alpha
\right\}  }\sum_{\left\{  \beta\right\}  }\langle0_{c}|\,W\,E_{\alpha^{\prime
}}E_{\alpha}^{\dagger}\,\,V^{\dagger}\,|0_{c}\rangle\,\\
&  L\left[  \left(  1,2\ldots x\right)  _{\alpha},\left(  1,2\ldots y\right)
_{\beta}\right]  \mathcal{\,A}_{\left\{  \beta\right\}  }K\left[  \left(
1,2\ldots x^{\prime}\right)  _{\alpha^{\prime}},\left(  1,2\ldots y\right)
_{\beta}\right] \, .
\end{align*}

Here $\mathcal{A}_{\left\{  \beta\right\}  }$ denotes a complete
anti-symmetrization over all indexes in the symbolic set $\left\{  \beta\right\}$. The result may be
proven by noticing that the complete contraction between the products
$C_{\beta^{\prime}}^{\dagger}C_{\beta}$ may be expressed as%
\[
C_{\beta^{\prime}}^{\dagger}C_{\beta}=\det\left|
\begin{array}
[c]{cccc}%
\delta_{11^{\prime}} & \delta_{12^{\prime}} & \cdots & \delta_{1x^{\prime}}\\
\delta_{21^{\prime}} & \delta_{22^{\prime}} & \cdots & \delta_{2x^{\prime}}\\
\vdots & \vdots & \ddots & \vdots\\
\delta_{x1^{\prime}} & \delta_{x2^{\prime}} & \cdots & \delta_{xx^{\prime}}%
\end{array}
\right| \, .
\]

The result of computing the remaining product
$\langle0_{c}|\,W\,E_{\alpha^{\prime}}E_{\alpha}^{\dagger}\,\,V^{\dagger}\,|0_{c}\rangle$
depends on whether we have the valence operators present inside the $W$ and $V^\dagger$ groups;
we treat three separate cases below.

\subsubsection{Both valence operators are absent}
In this case, both objects, $L$ and $K$, necessarily depend on their respective valence indexes and we emphasize
these dependencies by super-scripts $L^{v}$ and $K^{w}$. Also   $x=x^{\prime}$
and (similar to full contraction for core orbitals)
\[
\langle\Psi_{w}|\Phi_{v}\rangle_{\mathrm{val}}=\sum_{\left\{  \alpha\right\}
}\sum_{\left\{  \beta\right\}  }\,L^{v}\left[  \left(  1,2\ldots x\right)
_{\alpha},\left(  1,2\ldots y\right)  _{\beta}\right]  \mathcal{\,A}_{\left\{
\alpha\right\}  }\mathcal{\,A}_{\left\{  \beta\right\}  }K^{w}\left[  \left(
1,2\ldots x\right)  _{\alpha},\left(  1,2\ldots y\right)  _{\beta}\right] \, .
\]
Such a combination produces $x!y!$ diagrams.

\subsubsection{Only one valence operator is present}

Consider first the case when only the operator $a_{v}^{\dagger}$ is present, then
the object $K$ depends on the index $w$; we denote this dependence as $K^{w}$.
Also $x^{\prime}=x+1$. By appending $a_{v}^{\dagger}$ at the end of the $E_{\alpha}^{\dagger
}$ string, we reduce the treatment of contractions to the preceding case.
\[
\langle\Psi_{w}|\Phi_{v}\rangle=\left(  -1\right)  ^{\,y}\sum_{\left\{
\alpha^{\prime}\right\}  ,\left\{  \alpha\right\}  }\sum_{\left\{
\beta\right\}  }\langle0_{c}|\,\,E_{\alpha^{\prime}}E_{\alpha}^{\dagger
}\,\,a_{v}^{\dagger}|0_{c}\rangle\,L\left[  \left(  1,2\ldots x\right)
_{\alpha},\left(  1\ldots y\right)  _{\beta}\right]  \mathcal{\,A}_{\left\{
\beta\right\}  }K^{w}\left[  \left(  1^{\prime},2^{\prime}\ldots x^{\prime
}\right)  _{\alpha^{\prime}},\left(  1\ldots y\right)  _{\beta}\right] \, ,
\]%
\[
\langle\Psi_{w}|\Phi_{v}\rangle_{v}=\left(  -1\right)  ^{y}\sum_{\left\{
\alpha\right\}  }\sum_{\left\{  \beta\right\}  }L\left[  \left(  1,2\ldots
x\right)  _{\alpha},\left(  1\ldots y\right)  _{\beta}\right]  \mathcal{A}%
_{\left\{  \alpha\right\}  }\mathcal{A}_{\left\{  \beta\right\}  }K^{w}\left[
\left(  1,2\ldots x,v\right)  _{\alpha},\left(  1\ldots y\right)  _{\beta
}\right] \, .
\]

If only $a_{w}$ is present, then
\[
\langle\Psi_{w}|\Phi_{v}\rangle_{w}=\left(  -1\right)  ^{y}\sum_{\left\{
\alpha^{\prime}\right\}  }\sum_{\left\{  \beta\right\}  }K\left[  \left(
1^{\prime},2^{\prime}\ldots x^{\prime}\right)  _{\alpha},\left(  1\ldots
y\right)  _{\beta}\right]  \mathcal{A}_{\left\{  \alpha\right\}  }%
\mathcal{A}_{\left\{  \beta\right\}  }L^{v}\left[  \left(  1^{\prime
},2^{\prime}\ldots x^{\prime},w\right)  _{\alpha},\left(  1\ldots y\right)
_{\beta}\right] \, .
\]

\subsubsection{Both valence operators are present}
If both valence operators are present, i.e., $W=a_{w}$ and $V^{\dagger}
=a_{v}^{\dagger}$, then $x=x^{\prime}$. The product $\langle0_{c}%
|\,a_{w}E_{\alpha^{\prime}}E_{\alpha}^{\dagger}\,\,a_{v}^{\dagger}%
\,|0_{c}\rangle$ may be broken into two contributions.

(i) Contraction between $a_{w}$ and $a_{v}^{\dagger}$ leads to a core contribution
\begin{equation}
\langle\Psi_{w}|\Phi_{v}\rangle_{c}^{x=x^{\prime}}=\delta_{wv}\sum_{\left\{
\alpha\right\}  }\sum_{\left\{  \beta\right\}  }\,L\left[  \left(  1,\ldots
x\right)  _{\alpha},\left(  1,\ldots y\right)  _{\beta}\right]  \mathcal{\,A}%
_{\left\{  \alpha\right\}  }\mathcal{\,A}_{\left\{  \beta\right\}  }K\left[
\left(  1,\ldots x\right)  _{\alpha},\left(  1,\ldots y\right)  _{\beta
}\right] \, .
\label{Eq:Zcore}
\end{equation}
Notice that this contribution vanishes for $w\not=v$ and, moreover, it does not vanish only in a very special case of true scalar operator $Z$.

(ii) Simultaneous contractions between $a_{w}$ and an operator in $E_{\alpha}^{\dagger
}$, $a_{v}^{\dagger}$ with an operator in $E_{\alpha^{\prime}}$ and residual
contractions lead to $\left(  y!\,x!\right)  \,x$ contributions
\[
\langle\Psi_{w}|\Phi_{v}\rangle_{\mathrm{val}}=-\sum_{\left\{  \alpha\right\}
}\sum_{\left\{  \beta\right\}  }\underset{\xi\in\left\{  \alpha\right\}
}{{\LARGE S}}\,L\left[  \left(  1,\ldots x\right)  _{\alpha},\left(  1,\ldots
y\right)  _{\beta}\right]  _{\xi\rightarrow w}\mathcal{\,A}_{\left\{
\alpha\right\}  }\mathcal{\,A}_{\left\{  \beta\right\}  }K\left[  \left(
1,2\ldots x\right)  _{\alpha},\left(  1,\ldots y\right)  _{\beta}\right]
_{\xi\rightarrow v}
\]
Here $\underset{\xi\in\left\{  \alpha\right\}  }{{\LARGE S}}$ represents a
summation over all possible simultaneous replacement of index $\xi$ by $w$ in
$L$ and by $v$ in $K$.

\section{Counting diagrams and summary}
\label{Sec:Count}
The present paper provides symbolic prescriptions to aid in deriving
many-body diagrams for mono-valent systems in an arbitrary order of MBPT.
Based on the derived rules, the author has developed a Mathematica package,
which is made available through the author's website~\cite{APDweb}.
The rules have been tested by recovering known results for matrix elements
and energies through the third order~\cite{BluGuoJoh87}. Notice that
the package and the theorems have been already used in deriving
matrix elements through the fourth order of MBPT~\cite{DerEmm02,CanDer04}.

The derived rules are certainly not as mnemonically elegant as
the original Wick's theorem; this is a reflection of the fact that the Wick theorem
is formulated  in terms of the pair-wise contractions between the operators,
while our rules provide an explicit (yet general) answer that
avoids a multitude of elemental pair-wise contractions.  In this sense, the derived
theorems may be called ``post-Wick'' theorems.

As an illustration, we derive the MBPT diagrams for matrix elements of a one-body operator
between two distinct states $w$ and $v$ of a mono-valent atom.
We run the package and obtain  analytical expressions for the diagrams.
The expressions are too lengthy to be  presented here (see, however, a partial list of fourth-order diagrams in Ref.~\cite{DerEmm02}); instead we count the number of the resulting diagrams.
The results are compiled in Table~\ref{Tab:DiagCount}. The counts do not include  the normalization
term of Eq.(\ref{Eq:Zn}) and the core contribution (\ref{Eq:Zcore}).

\begin{table}[h]
\caption{Complexity of MBPT for mono-valent systems in the frozen-core approximation.
We list numbers of diagrams in the $n^\mathrm{th}$ order of MBPT for wave-functions and matrix elements of
a one-particle operator. There are two counts for $Z_{wv}^{(n)}$ in the format $n_1/n_2$. $n_1$ is a number of terms in  a maximally simplified expression (where the Coulomb integrals $g_{ijkl}$ were combined into the anti-symmetric combinations $\tilde{g}_{ijkl}$). $n_2$ is the full number of the conventional Brueckner-Goldstone diagrams including
exchange diagrams. }
\label{Tab:DiagCount}
\begin{ruledtabular}
\begin{tabular}{crc}
\multicolumn{1}{c}{Order $n$} &
\multicolumn{1}{c}{$|\Psi_v^{(n)}\rangle$} &
\multicolumn{1}{c}{$Z_{wv}^{(n)}$}
 \\
\hline
0 &  1 & - \\
1 &  2 & 1/1 \\
2 & 20 & 2/4 \\
3 & 561 & 30/84 \\
4 &26700    & 552/3072 \\
\end{tabular}
\end{ruledtabular}
\end{table}

As discussed, due to our short-cutting the expensive elemental pair-wise contractions, the advantage of the present formulation becomes
most pronounced in high orders of perturbation theory,
as both the number of resulting terms rapidly grows (see Table~\ref{Tab:DiagCount}) and  the   length of operator
strings entering $|\Psi^{(n)}\rangle$ increases with $n$.
In addition, from a practical standpoint, in symbolic algebra implementations,
the next step involves combining similar terms by  pattern matching.
This is a computationally expensive search operation. By contrast to the
directly applied Wick's expansion method, our derived theorems
already yield simplified results further speeding up the symbolic evaluations.

\acknowledgements

This work was supported in part by the  U.S. National Science Foundation.


\end{document}